\documentclass[aps,amssymb,superscriptaddress,footinbib,twocolumn]{revtex4}
\usepackage{mathtools}
\usepackage{amsfonts}
\usepackage{latexsym}
\usepackage{relsize}
\usepackage{euscript}
\usepackage{amssymb}
\usepackage{graphicx}
\usepackage{amsmath}
\usepackage{amsbsy}
\usepackage{amsthm}
\usepackage{bbm}
\usepackage{bm}
\usepackage{epsfig}
\usepackage{epstopdf}
\usepackage{dsfont}
\usepackage{color}
\usepackage[colorlinks]{hyperref}
\usepackage[figure,table]{hypcap}
\usepackage[symbol]{footmisc}
\usepackage{enumerate}
\hypersetup{
	bookmarksnumbered,
	pdfstartview={FitH},
	citecolor={darkgreen},
	linkcolor={darkblue},
	urlcolor={darkblue},
	pdfpagemode={UseOutlines}}
\definecolor{darkgreen}{RGB}{50,190,50}
\definecolor{darkblue}{RGB}{0,0,190}
\definecolor{darkred}{RGB}{238,0,0}
\usepackage{soul}
\DeclareMathOperator\arctanh{arctanh}

\newcommand{\tr}{\operatorname{tr}}

\newcommand{\be}{\begin{equation}}
\newcommand{\ee}{\end{equation}}
\newcommand{\beq}{\begin{eqnarray}}
\newcommand{\eeq}{\end{eqnarray}}

\begin{document}

\title{Quantum metrology for relativistic quantum fields}

\author{Mehdi Ahmadi}
\email{mehdi.ahmadi@nottingham.ac.uk}
\affiliation{School of Mathematical Sciences, University of Nottingham, University Park,
Nottingham NG7 2RD, United Kingdom}
\author{David Edward Bruschi}\thanks{Current affiliation: Racah Institute of Physics and Quantum Information Science Centre, the Hebrew University of Jerusalem, 91904 Jerusalem, Israel}
\affiliation{School of Electronic and Electrical Engineering, University of Leeds, Woodhouse Lane,  Leeds, LS2 9JT,  United Kingdom}
\author{Ivette Fuentes}\thanks{Previously known as Fuentes-Guridi and Fuentes-Schuller.}
\affiliation{School of Mathematical Sciences, University of Nottingham, University Park,
Nottingham NG7 2RD, United Kingdom}

\begin{abstract}
In quantum metrology quantum properties such as squeezing and entanglement are exploited in the design of a new generation of clocks, sensors and other measurement devices that can outperform their classical counterparts. Applications of great technological relevance lie in the precise measurement of parameters which play a central role in relativity, such as proper accelerations, relative distances, time and gravitational field strengths. In this paper we generalise recently introduced techniques to estimate physical quantities within quantum field theory in flat and curved space-time. We consider a bosonic quantum field that undergoes a generic transformation, which encodes the parameter to be estimated. We present analytical formulas for optimal precision bounds on the estimation of small parameters in terms of Bogoliubov coefficients for single mode and two-mode Gaussian channels. \end{abstract}

\maketitle

\section{Introduction}

 Quantum metrology provides techniques to enhance the precision of measurements of physical quantities by exploiting quantum properties such as squeezing and entanglement. Such techniques are being employed to design a new generation of quantum measurement technologies such as quantum clocks  and sensors. Impressively, the quantum era is now reaching relativistic regimes. Table-top experiments demonstrate relativistic effects in quantum fields \cite{casimirwilson} and long range quantum experiments \cite{zeilingerteleport} will soon reach regimes where relativity kicks in  \cite{{rideout}, {SchillerEtalSpaceOpticalClock2012}}. It is therefore of great interest to develop quantum metrology techniques to measure physical quantities that play a role in relativity. These can lead to the measurement of gravitational waves and applications in relativistic geodesy, positioning, sensing and navigation. 
 
 Recently, techniques that apply quantum metrology to quantum field theory in curved and flat space-time have been developed \cite{aspachs,RQM,downes}. Quantum field theory allows one to incorporate relativistic effects at the regimes that are being reached by cutting-edge quantum experiments. 
 The application of quantum metrology to quantum field theory can in principle produce technologies that outperform (non-relativistic) quantum estimation of Newtonian gravitational parameters \cite{RQM}. Indeed, it was shown that relativistic effects can be exploited to improve measurement technologies \cite{RQM}. 
 
In this paper we generalise metrology techniques to estimate physical quantities that play a role in relativistic quantum field theory.  Previous work aligned with this spirit showed that entanglement can be used to determine the expansion rate of the universe~\cite{fredericivy} and that phase estimation techniques could be employed to measure the Unruh effect at accelerations that are within experimental reach~\cite{aspachs,HoslerKok2013}.  The limitations imposed by the quantum uncertainty principle in the measurement of space-time parameters were investigated within locally covariant quantum field theory ~\cite{downes} and non-relativitic quantum mechanics \cite{SaleckerWigner}.  Reference \cite{RQM} shows that quantum metrology techniques employing the covariance matrix formalism are ideally suited to estimate parameters in quantum field theory. The techniques were applied to the estimation of parameters of a field contained in a moving cavity undergoing arbitrary motion.  This led to the design of an accelerometer that can improve by two orders of magnitude the state-of-the-art optimal measurement precision.  Its important to point out that the equations presented in \cite{RQM} are only directly applicable to the example of the moving cavity. 
 
In this paper we generalize the techniques introduced in \cite{RQM}  by considering a bosonic quantum field that undergoes a generic transformation, which encodes the parameter to be estimated. Such transformation can involve, for example, space-time dynamics such as the expansion of the Universe,  a  general change of an observer's reference frame, as well as the example of a cavity undergoing non-uniform motion. In the case that the transformation admits a series expansion around the parameter to be estimated, we present general expressions for optimal precision bounds in terms of Bogoliubov coefficients.  Our analysis is restricted to single mode and two-mode Gaussian states for which elegant and simple formulas can be provided.  

The structure of this paper is as follows: in the next section we review notions from quantum field theory showing how the state of the field and its transformations can be described in the covariance matrix formalism. In section III, we will review quantum metrology techniques where the quantum Fisher information provides optimal bounds on measurement precisions. In section IV we generalise techniques for relativistic quantum metrology by providing analytical expressions for the quantum Fisher information in terms of the Bogoliubov coefficients of general transformations. In the last section, we compare single-mode and two-mode channels showing, in particular, that two-mode channels improve measurement precision in the example of a non-uniformly moving cavity.  

\section{Quantum field theory in curved space-time and the covariance matrix formalism }

We start by describing the states of a bosonic quantum field that undergoes a transformation which encodes the parameter we want to estimate. We consider a real massless scalar quantum field $\Phi$ in a space-time with metric $g_{\mu\nu}$. The signature of the metric is $(-,+,+,+)$. The field obeys the Klein-Gordon equation  $\square\Phi=0$ where the d'Alambertian takes the form $\square:=(\sqrt{-g})^{-1}\partial_{\mu}\sqrt{-g}g^{\mu\nu}\partial_{\nu}$. Here  $g=\text{det}(g_{\mu\nu})$ is the determinant of the metric. The field can be expanded in terms of a discrete set of modes \cite{footnote},
\begin{equation}
\Phi=\sum_n\left[\phi_{n}a_n+\text{h.c.}\right],
\end{equation} where the creation and annihilation operators, $a_n^{\dagger}$ and $a_n$, satisfy the canonical bosonic commutation relations $[a_{m},a^{\dag}_{n}]=\delta_{mn}$.  The modes $\{\phi_{n}|n=1,2,3,...\}$ are solutions to the Klein-Gordon equation and form a complete set of orthonormal modes with respect to the inner product, denoted by $(\,\cdot\,,\,\cdot\,)$~\cite{BirrellDavies:QFbook}. 
The vacuum state $|0\rangle$ of the field is defined as the state that is annihilated by the operators $a_{n}$ for all~$n$, i.e., $a_{n} |0\rangle=0$.  

It is important to note that the mode decomposition is not unique, the field can be decomposed in terms of a new set of modes which we denote $\tilde{\phi}_{n}$.
For example, a coordinate transformation between different observers results in a Bogoliubov transformation between the modes $\phi_{n}$ and mode solutions $\tilde{\phi}_{n}$ in the new coordinate system. 

Indeed a Bogoliubov transformation is the most general linear transformation between two sets of field modes ${\phi}_{n}$ and $\tilde{\phi}_{n}$. The transformation between the corresponding operators $a_{m}$ and $\tilde{a}_{n}$ is given by
\be\label{Bogoliubov transformation}
\tilde{a}_{m}=\sum_{n} \bigl(\alpha^{*}_{mn}a_{n}-\beta^{*}_{mn}a^{\dag}_{n}\bigr)\,,
\ee
where $\alpha_{mn}:=(\tilde{\phi}_{m},\phi_{n})$ and $\beta_{mn}:=(\tilde{\phi}_{n},\phi^{*}_{m})$ are the Bogoliubov coefficients.  The operators $\tilde{a}_{n}$ define the vacuum state $|\tilde{0}\rangle$ in the new basis through the condition $\tilde{a}_{n} |\tilde{0}\rangle=0$. Paradigmatic examples are the Bogoliubov transformations
 between inertial and uniformly accelerated observers in flat space-time, between Kruskal and Schwarzschild observers in a black-hole space-time and between observers at past and future infinity in an expanding Universe \cite{BirrellDavies:QFbook}.

In quantum field theory different observers not always agree on the particle content of a state \cite{BirrellDavies:QFbook}. Note that $|\tilde{0}\rangle$ is annihilated by the initial field operators $a_n$ only if all coefficients $\beta_{mn}$ are zero. Therefore, $\beta_{mn}\neq0$ is an indication of the particle creation. This occurs, for instance, in the Unruh effect where the inertial vacuum state is seen as a thermal state by uniformly accelerated observers~\cite{BirrellDavies:QFbook}.  
 An example of particular interest in this work is that of a cavity undergoing non-inertial motion \cite{alphacentauri,ourreview,nicoivy}. The vacuum state of an inertial cavity becomes populated after the cavity undergoes non-uniformly accelerated motion~\cite{casimirwilson}. As we mentioned before, the Bogoliubov coefficients depend on the parameter we want to estimate. Examples of such parameters are proper time, the expansion rate of the Universe, the mass of a black hole, the acceleration of a cavity, among others. 
 
Let us now describe the state of the field and its Bogoliubov transformations in the covariance matrix formalism. This formalism has been used to investigate entanglement in quantum field theory  ~\cite{givyericsson,nicoivy,gcqg} since it is applicable to systems consisting of a (discrete) infinite number of bosonic modes. The covariance matrix formalism enables elegant and simplified calculations to quantify bipartite and multipartite entanglement for Gaussian states~\cite{ourreview,gcqg}. In order to introduce this formalism we define the quadrature operators $X_{2n-1}=\frac{1}{\sqrt{2}}(a_{n}+a^{\dag}_{n})$ and $X_{2n}=\frac{1}{\sqrt{2}\,i}(a_{n}-a^{\dag}_{n})$, which correspond to the generalised position and momentum operators of the field. 
We will restrict our analysis to Gaussian states since they take a simple form in this formalism. These states are completely described by the expectation values of the quadratures $\langle X_{i}\rangle$ (also known as the first moments of the field) and the covariance matrix $\Sigma_{ij}=\langle X_{i} X_{j}+X_{j}X_{i}\rangle-2\langle X_{i}\rangle\langle X_{j}\rangle$. The unitary transformations in the Hilbert space that are generated by a quadratic Hamiltonian can be represented as a symplectic matrix $S$ in phase space. These transformations form the real symplectic group $Sp(2n,\mathds{R})
$, the group of real $(2n\times2n)$ matrices that leave the symplectic form $\Omega$ invariant, i.e., $S\Omega S^{T}=\Omega$, where $\Omega=\bigoplus_{k=1}^{n}\Omega_k$ and $\Omega_k=-i\sigma_y$ and $\sigma_y$ is one of the Pauli matrices. The time evolution of the field, as well as the Bogoliubov transformations, can be encoded in this structure. The symplectic matrix corresponding to the Bogoliubov transformation in Eq.~(\ref{Bogoliubov transformation}) can be written in terms of the Bogoliubov coefficients as
\be\label{Bogosymplectic}
S=\left(
  \begin{array}{cccc}
    \mathcal{M}_{11} & \mathcal{M}_{12} & \mathcal{M}_{13} & \cdots \\
    \mathcal{M}_{21} & \mathcal{M}_{22} & \mathcal{M}_{23} & \cdots \\
    \mathcal{M}_{31} & \mathcal{M}_{32} & \mathcal{M}_{33} & \cdots \\
    \vdots & \vdots & \vdots & \ddots
  \end{array}
\right)\,,
\ee
where the $\mathcal{M}_{mn}$ are the $2\times2$ matrices
\be\label{Mmatrices}
\mathcal{M}_{mn}=\left(
                   \begin{array}{cc}
                     \Re(\alpha_{mn}-\beta_{mn}) & \Im(\alpha_{mn}+\beta_{mn}) \\
                     -\Im(\alpha_{mn}-\beta_{mn}) & \Re(\alpha_{mn}+\beta_{mn})
                   \end{array}
                 \right)\,.
\ee
Here $\Re$ and $\Im$ denote the real and imaginary parts, respectively. The first moments and covariance matrix after a Bogoliubov transformation are given by $\langle \tilde{X} \rangle=S\langle X_0 \rangle$ and $\tilde{\Sigma}=S\Sigma_0 S^{T}$ respectively, where $\langle X_0 \rangle$ and $\Sigma_0$ encode the initial state of the field.  In this formalism it is relatively easy to apply metrology techniques to estimate physical quantities which are encoded in the Bogoliubov coefficients.  

\section{Quantum metrology} 
In this section we briefly review metrology techniques for Gaussian states. Quantum metrology deals with the estimation of quantities that do not correspond to an observable of the system \cite{advances}. Examples are temperature, time, acceleration, and coupling strengths.  The techniques assume that the state of a quantum system undergoes a transformation which encodes the parameter to be estimated. The main aim is to find the optimal estimation strategy, i.e. finding an optimal initial state and the set of measurements on the final state that will allow us to estimate the parameter with the highest possible precision. We are interested in estimating the precision in the case one measures only one or two modes in a Gaussian state. Note that symplectic transformations take Gaussian states into Gaussian states.

A parameter $\theta$ can be estimated with high accuracy when the states $\rho_{\theta}$ and $\rho_{\theta+d\theta}$, which differ from each other by an infinitesimal change $d\theta$, can be distinguished. The operational measure that quantifies the distinguishability of these two states is the Fisher information \cite{paris}. Let us suppose that an agent performs $N$ independent measurements to obtain an unbiased estimator $\hat{\theta}$ for the parameter $\theta$. The Fisher Information $F(\theta)$ tells us how much information can be extracted about the unknown parameter $\theta$. In other words it gives us a lower bound to the  mean-square error in the estimation of $\theta$ via the classical Cram$\mathrm{\acute{e}}$r-Rao inequality~\cite{Cramer:Methods1946}, i.e., $\langle (\Delta \hat{\theta})^{2}\rangle\geq\frac{1}{NF(\theta)}$, where $F(\theta)=\int\!d\lambda~p(\lambda|\theta) (d\,\ln [p(\lambda|\theta)]/d\lambda)^{2}$ and $p(\lambda|\theta)$ is the likelihood function with respect to a chosen positive operator valued measurement (POVM)~$\{\hat{O}_{\lambda}\}$ with $\sum_{\lambda}\hat{O}_{\lambda}=\mathds{1}$\,. 
Optimizing over all the possible quantum measurements provides an even stronger lower bound~\cite{BraunsteinCaves1994}, i.e.,
\be\label{Cramer-Rao}
N\langle (\Delta \hat{\theta})^{2}\rangle\geq\frac{1}{F(\theta)}\geq \frac{1}{H(\theta)},
\ee
where $H(\theta)$ is the quantum Fisher information (QFI). This quantity is obtained by determining the eigenstates of the symmetric logarithmic derivative $\Lambda_{\rho_\theta}$ defined by  $2\frac{d\rho_{\theta}}{d\theta}=\Lambda_{\rho_{\theta}} \rho_{\theta}+\rho_{\theta} \Lambda_{\rho_{\theta}}\,$. Alternatively, the QFI can be related to the Uhlmann fidelity~$\mathcal{F}$ of the two states $\rho_{\theta}$ and $\rho_{\theta+d\theta}$ through
 \begin{equation}
 H(\theta)=\frac{8\big(1-\sqrt{\mathcal{F}(\rho_{\theta},\rho_{\theta+d\theta})}\big)}{d\theta^{2}}, \label{quantumfishinfo}
\end{equation}
where $\mathcal{F}(\rho_1,\rho_2)=(\tr\sqrt{\sqrt{\rho_1}\rho_2\sqrt{\rho_1}})^{2}\,$. The optimal masurements for which the quantum Cram$\mathrm{\acute{e}}$r-Rao bound~(\ref{Cramer-Rao}) becomes asymptotically tight can be computed from $\Lambda_{\rho_\theta}$ \cite{monras}. Unfortunately, these optimal measurements are usually not easily implementable in the laboratory. Nevertheless, in typical problems involving optimal implementations one can devise suboptimal strategies involving feasible measurements such as homodyne or heterodyne detection, see, e.g.,~\cite{advances}. In this paper we are interested in comparing the quantum Cram$\mathrm{\acute{e}}$r-Rao bound for different initial states.

Expressions for the fidelity for one and two mode states have been previously obtained. 
The fidelity between two generally mixed single mode Gaussian states $\sigma$ and $\sigma'$ is given by~\cite{MarianMarian}
\begin{equation}
\label{SinglemodeMarianFidelity}
F(\sigma,\sigma')=\frac{1}{\sqrt{\Lambda+\Delta}-\sqrt{\Lambda}} \exp[-\langle \Delta X\rangle^{T} A^{-1}\langle \Delta X\rangle],
\end{equation}
where 
\begin{align*}
A =&\,\sigma+\sigma',\\
\langle \Delta X\rangle =&\,\langle X\rangle_{\sigma'}-\langle X\rangle_{\sigma},\\
\Delta =&\,\frac{1}{4}\det(A),\\
\Lambda =&\, \frac{1}{4} \det(\sigma+ i \Omega )\det(\sigma'+ i \Omega ).
\end{align*}
Note that we follow the conventions used in~\cite{ourreview,nicoivy} for the normalization of the covariance matrix, which differ from other conventions~\cite{MarianMarian}.
 
Now let $\sigma$ and $\sigma'$ be two-mode Gaussian states with zero initial first moments. The Fidelity between them is given by
\cite{MarianMarian}
\begin{eqnarray}\label{FidelityMarianTwomodes}
    \mathcal{F}(\sigma,\sigma'),=\frac{\exp[-\langle \Delta X\rangle^{T} A^{-1}\langle \Delta X\rangle]}{\sqrt{\Lambda}+\sqrt{\Gamma}-\sqrt{(\sqrt{\Lambda}+\sqrt{\Gamma})^2-\Delta}},\label{two:mode:fidelity}
\end{eqnarray}
where
\begin{align}
\Gamma =&\,\frac{1}{16}\text{det}(i\mathbf{\Omega}\sigma i\mathbf{\Omega}\sigma'+\mathds{1})\nonumber\\
\Lambda =&\,\frac{1}{16}\text{det}(i\mathbf{\Omega}\sigma+\mathds{1})\text{det}(i\mathbf{\Omega}\sigma'+\mathds{1})\nonumber\\
\Delta =&\,\frac{1}{16}\text{det}(A)\nonumber\\
A=&\, \sigma+\sigma',
\end{align}
and $\mathds{1}$ is the identity matrix.
 
\section{Relativistic quantum metrology}

We now develop the main formalism contained in this work. We consider a bosonic quantum field which undergoes a general Bogoliubov transformation that depends on a dimensionless parameter~$\theta$, where $\theta$ is the parameter we want to estimate. We assume that the Bogoliubov coefficients which relate the initial and final state of the field have a series expansion in terms of $\theta$, 
\begin{subequations}
\label{Maclaurin expansion}
\begin{align}
\alpha_{mn} &=\,    \alpha^{\raisebox{0.5pt}{\tiny{$\,(0)$}}}_{mn}\,+\,
                    \alpha^{\raisebox{0.5pt}{\tiny{$\,(1)$}}}_{mn}\,\theta\,+\,
                    \alpha^{\raisebox{0.5pt}{\tiny{$\,(2)$}}}_{mn}\,\theta^{2}\,+\,O(\theta^{3})\,,
                    \label{Maclaurin expansion alphas}\\
\beta_{mn}  &=\,    \beta^{\raisebox{0.5pt}{\tiny{$\,(1)$}}}_{mn}\,\theta\,+\,
                    \beta^{\raisebox{0.5pt}{\tiny{$\,(2)$}}}_{mn}\,\theta^{2}\,+\,O(\theta^{3}).
                    \label{Maclaurin expansion betas}
\end{align}
\end{subequations}
This implies that our formalism applies only to small parameters $\theta$. Such assumption enables us to provide analytical expressions for the quantum Fisher information in terms of general Bogoliubov coefficients. 
A comment about the small parameter $\theta$ is in order.  Note that $\theta$ will generally be a function of a dimension-full parameter we ultimately want to estimate which needs not to be small in absolute terms. As an example consider $\theta=v/c$, where $v$ is the velocity of the system and $c$ is the speed of light. In this case, our perturbative treatment allows for the estimation of velocities as large as $v\sim0.1 c$.

To find the leading order term in $\theta$ of the quantum Fisher information we need to compute the second order term in the expansion of the fidelity, i.e. the coefficient of $d\theta^2$.  Equations (\ref{Maclaurin expansion}) lead to an expansion for the symplectic transformation 
\begin{equation}
\mathbf{S}(\theta)=\mathbf{S}^{\raisebox{0.5pt}{\tiny{$\,(0)$}}}+\mathbf{S}^{\raisebox{0.5pt}{\tiny{$\,(1)$}}}\theta+\mathbf{S}^{\raisebox{0.5pt}{\tiny{$\,(2)$}}}\theta^2+O(\theta^3).
\end{equation}
Therefore, the first moments and the covarinace matrix can be written as
\begin{align}\label{FME}
\langle X \rangle_{\theta}&=\mathbf{S}(\theta) \langle X_0 \rangle\nonumber\\
&=\langle X \rangle^{\raisebox{0.5pt}{\tiny{$\,(0)$}}}+\langle X \rangle^{\raisebox{0.5pt}{\tiny{$\,(1)$}}}\theta+\langle X \rangle^{\raisebox{0.5pt}{\tiny{$\,(2)$}}}\theta^2+O(\theta^3)
\end{align}
and
 \begin{equation}\label{CM}
\sigma(\theta)=\sigma^{\raisebox{0.5pt}{\tiny{$\,(0)$}}}+\sigma^{\raisebox{0.5pt}{\tiny{$\,(1)$}}}\theta+\sigma^{\raisebox{0.5pt}{\tiny{$\,(2)$}}}\theta^2+O(\theta^3).
  \end{equation}
In general, it is of interest to analyze the evolution of a finite set of modes of the field. Using Willamson's decomposition \cite{Weedbrook}, we can write the marginal covariance matrix of the $N$-modes as $\sigma(\theta)=\mathbf{s}(\theta)\sigma_{\oplus}(\theta)\mathbf{s}^{\dagger}(\theta)$, where $\mathbf{s}$ is a $2N\times2N$ symplectic transformation with the expansion 
\begin{equation}
\mathbf{s}(\theta)=\mathbf{s}^{\raisebox{0.5pt}{\tiny{$\,(0)$}}}+\mathbf{s}^{\raisebox{0.5pt}{\tiny{$\,(1)$}}}\theta+\mathbf{s}^{\raisebox{0.5pt}{\tiny{$\,(2)$}}}\theta^2+O(\theta^3),
\end{equation}
 and $\sigma_{\oplus}(\theta)=\text{diag}(\nu_{1}(\theta),\nu_{2}(\theta),\nu_{1}(\theta),\nu_{2}(\theta))$ is known as the symplectic form of $\mathbf{s}(\theta)$. The functions $\nu_{i}(\theta)$ are called the ``symplectic eigenvalues'' of $\mathbf{s}(\theta)$. They are simply the eigenvalues of the matrix $|i\mathbf{\Omega}\sigma(\theta)|$. In general $\nu_{i}\geq 1$, where the equality holds for pure states.
 
In this paper we consider states that are initially pure and that remain pure to lowest order in the parameter $\theta$ i.e., $\det [\sigma(\theta)]=1+O(\theta)$. This is means that the Bogoliubov transformations do not entangle modes to zero order. This requirement, together with the unitarity condition of the Bogoliubov transformations, implies that the $\alpha$-coefficients to zero order are $\alpha^{\raisebox{0.5pt}{\tiny{$\,(0)$}}}_{mn}=G_{n}\delta_{mn}$, where $G_{n}=\exp(i\phi_{n})$. This also implies that the expansion of the symplectic eigenvalues $\nu_{i}(\theta)$ in terms of $\theta$ is 
\begin{equation}
\nu_{i}(\theta)=1+\nu^{\raisebox{0.5pt}{\tiny{$\,(1)$}}}_{i}\theta+\nu^{\raisebox{0.5pt}{\tiny{$\,(2)$}}}_{i}\theta^2+O(\theta^3).
\end{equation} 
Imposing the necessary constraints $\mathcal{F}(\sigma(\theta),\sigma(\theta))=1$ and \cite{BraunsteinCaves}
\begin{equation}
\frac{\partial \mathcal{F}(\sigma(\theta),\sigma(\theta+d\theta))}{\partial d\theta}\big|_{d\theta=0}=0,
\end{equation}
allows us to write the expansion of fidelity as 
\begin{equation}\label{two-mode general formula}
\mathcal{F}(\sigma(\theta),\sigma(\theta+d\theta))= 1-\mathcal{F}^{(2)}\frac{d\theta^2}{2}+O(\theta d\theta^2+\theta^2 d\theta).
\end{equation}
We find that the second order term in the fidelity has two contributions $\mathcal{F}^{\raisebox{0.5pt}{\tiny{$\,(2)$}}}=\mathcal{E}^{\raisebox{0.5pt}{\tiny{$\,(2)$}}}+\mathcal{C}^{\raisebox{0.5pt}{\tiny{$\,(2)$}}}$. The first contribution comes form the expansion of the exponential term in the fidelity formulas (\ref{SinglemodeMarianFidelity}) and \eqref{FidelityMarianTwomodes} which is given by
\begin{eqnarray}\label{Displacementfidelity}
e^{-\langle \Delta X\rangle^{T} A^{-1}\langle \Delta X\rangle}&=&1-\mathcal{E}^{\raisebox{0.5pt}{\tiny{$\,(2)$}}}d\theta^{2}\nonumber\\
&=&1-{\langle X \rangle^{\raisebox{0.5pt}{\tiny{$\,(1)$}}}}^{T} {A^{-1}}^{\raisebox{0.5pt}{\tiny{$\,(0)$}}}\langle X \rangle^{\raisebox{0.5pt}{\tiny{$\,(1)$}}}d\theta^{2},\nonumber\\
\end{eqnarray}
where ${A^{-1}}^{\raisebox{0.5pt}{\tiny{$\,(0)$}}}={A^{{\raisebox{0.5pt}{\tiny{$\,(0)$}}}}}^{-1}$.  The second contribution $\mathcal{C}^{\raisebox{0.5pt}{\tiny{$\,(2)$}}}$ comes from the denominator in equations (\ref{SinglemodeMarianFidelity}) and \eqref{FidelityMarianTwomodes}. Having found the second order contributions to the fidelity we can write the QFI as 
\begin{equation}\label{QFI}
H=4(\mathcal{E}^{\raisebox{0.5pt}{\tiny{$\,(2)$}}}+\mathcal{C}^{\raisebox{0.5pt}{\tiny{$\,(2)$}}}). 
\end{equation}
In the following sections we compute $\mathcal{E}^{\raisebox{0.5pt}{\tiny{$\,(2)$}}}$ and $\mathcal{C}^{\raisebox{0.5pt}{\tiny{$\,(2)$}}}$ for single-mode and two-mode detection schemes.

\subsubsection{Single-mode detection}
We start by considering the scenario in which a single mode $k$ of the quantum field is detected to estimate the parameter $\theta$. 
 Therefore, we assume that all of the modes of the quantum field are initially in the vacuum state except for mode $k$ which is in a squeezed displaced vacuum, i.e.
$|\Psi_0\rangle=S_k(r)D_k(\delta) |0\rangle$, where $D_k(\delta)=\exp[\delta( a_k^{\dag}- a_k)]$ and $S_k(r)=\exp[\frac{r}{2}( a_k^{\dag^2}- a_k^{^2})]$ are the displacement and squeezing operators, respectively.  The parameters $\delta$ and $r$ determine the amount of displacement and squeezing in the state. 
 The corresponding reduced $2\times2$ covariance matrix of mode $k$ takes the form $\sigma_{0}=\text{diag}(e^{r}, e^{-r})$. The reduced covariance matrix of mode $k$ after the Bogoliubov transformation can be written as 
\begin{equation}\label{CMSM}
\sigma_{k}(\theta)=\mathcal{M}_{kk}(\theta) \sigma_{0} \mathcal{M}_{kk}^{T}(\theta)+\sum_{n\neq k}\mathcal{M}_{kn}(\theta)\mathcal{M}_{kn}^{T}(\theta),
\end{equation}
where the matrices $\mathcal{M}_{ij}(\theta)$ can be found using equations (\ref{Mmatrices}) and (\ref{Maclaurin expansion}).  

Let us first compute the contribution $\mathcal{E}^{\raisebox{0.5pt}{\tiny{$\,(2)$}}}$ to the fidelity. In order to do so we first need to compute the first moments of the quantum field after the Bogoliubov transfomation. The initial first moments of the mode $k$ of the quantum field can be written as $\langle X_{k}\rangle=(\sqrt{2}\delta,0)^{T}$. Now using the equation \eqref{FME} we can write the first contribution to the final first moments as 
\begin{equation}
\langle  X \rangle^{\raisebox{0.5pt}{\tiny{$\,(1)$}}}= \left(
  \begin{array}{c}
    \sqrt{2}\delta\Re[\alpha_{kk}^{\raisebox{0.5pt}{\tiny{$\,(1)$}}}-\beta_{kk}^{\raisebox{0.5pt}{\tiny{$\,(1)$}}}]\\
  - \sqrt{2}\delta\Im[\alpha_{kk}^{\raisebox{0.5pt}{\tiny{$\,(1)$}}}-\beta_{kk}^{\raisebox{0.5pt}{\tiny{$\,(1)$}}}]\\ 
  \end{array}
\right)\,,
\end{equation}
 Using \eqref{Displacementfidelity} we find that the second order contribution $\mathcal{E}^{\raisebox{0.5pt}{\tiny{$\,(2)$}}}$  to the fidelity is given by
\begin{eqnarray}
\mathcal{E}_{1}^{\raisebox{0.5pt}{\tiny{$\,(2)$}}}&=&2\delta^2\big( \big|\alpha_{kk}^{\raisebox{0.5pt}{\tiny{$\,(1)$}}}-\beta_{kk}^{\raisebox{0.5pt}{\tiny{$\,(1)$}}}\big|^2\cosh r\nonumber\\
&+&\Re[(\alpha_{kk}^{\raisebox{0.5pt}{\tiny{$\,(1)$}}}-\beta_{kk}^{\raisebox{0.5pt}{\tiny{$\,(1)$}}})^2 {G^{*}_{k}}^{2}]\sinh r\big).
\end{eqnarray}
Note that this term depends both on the amount of squeezing $r$ and the amount of displacement $\delta$.

In order to obtain the contribution $\mathcal{C}^{\raisebox{0.5pt}{\tiny{$\,(2)$}}}$ to the fidelity, we need to find the perturbative expansions of $\Lambda$ and $\Delta$. Using the equation \eqref{SinglemodeMarianFidelity} we find  
 \begin{eqnarray}
\Delta(\theta, \theta+d \theta)&=&1+ \Delta^{\raisebox{0.5pt}{\tiny{$\,(2)$}}} \frac{d\theta^2}{2}+\Delta^{\raisebox{0.5pt}{\tiny{$\,(2)$}}}(\theta+d\theta)d\theta+ O(\theta^2 d\theta^2)\nonumber\\
\Lambda(\theta, \theta+d \theta)&=& \frac{1}{4}( \Delta^{\raisebox{0.5pt}{\tiny{$\,(2)$}}} \theta (\theta+d\theta))^2+O(\theta^3 (\theta+d\theta)^3 ),
\end{eqnarray} 
where 
\begin{eqnarray}\label{Delta2}
\Delta^{\raisebox{0.5pt}{\tiny{$\,(2)$}}}&=& 2 (\sigma^{\raisebox{0.5pt}{\tiny{$\,(0)$}}}_{11}\sigma^{\raisebox{0.5pt}{\tiny{$\,(2)$}}}_{22}+\sigma^{\raisebox{0.5pt}{\tiny{$\,(2)$}}}_{11}\sigma^{\raisebox{0.5pt}{\tiny{$\,(0)$}}}_{22}-2\sigma^{\raisebox{0.5pt}{\tiny{$\,(0)$}}}_{12}\sigma^{\raisebox{0.5pt}{\tiny{$\,(2)$}}}_{12})\nonumber\\
&+&\frac{1}{2}(\sigma^{\raisebox{0.5pt}{\tiny{$\,(1)$}}}_{11}\sigma^{\raisebox{0.5pt}{\tiny{$\,(1)$}}}_{22}-2(\sigma^{\raisebox{0.5pt}{\tiny{$\,(1)$}}}_{12})^2),
\end{eqnarray}
and $\sigma^{\raisebox{0.5pt}{\tiny{$\,(n)$}}}_{ij}$s are the elements of $n$th order matrices in \eqref{CM}.
After some algebra we find the second contribution $\mathcal{C}^{\raisebox{0.5pt}{\tiny{$\,(2)$}}}$ to the fidelity formula  \eqref{SinglemodeMarianFidelity} as, 
\begin{eqnarray}\label{QFIsinglemodegenral}
\mathcal{C}_{1}^{\raisebox{0.5pt}{\tiny{$\,(2)$}}}&=& \frac{\Delta^{\raisebox{0.5pt}{\tiny{$\,(2)$}}}}{4}\nonumber\\
&=&[(f_{\alpha}^{k}+f_{\beta}^{k} )\cosh(r)-\sum_{n \neq k}\Re[ \alpha^{\raisebox{0.5pt}{\tiny{$\,(1)$}}}_{nk} \beta^{\raisebox{0.5pt}{\tiny{$\,(1)$}}*}_{nk}] \sinh(r)\nonumber\\
&-&\frac{1}{4}\big(\Re[ \alpha^{\raisebox{0.5pt}{\tiny{$\,(1)$}}}_{kk} \beta^{\raisebox{0.5pt}{\tiny{$\,(1)$}}*}_{kk}] \sinh(2r)+
|\alpha^{\raisebox{0.5pt}{\tiny{$\,(1)$}}}_{kk} |^2 \cosh^2(r)\nonumber\\
&+&\frac{1}{2}(|\beta^{\raisebox{0.5pt}{\tiny{$\,(1)$}}}_{kk}|^2\sin^2(\phi_{k})+ \frac{1}{2}\Im[\beta^{\raisebox{0.5pt}{\tiny{$\,(1)$}}}_{kk}]\Re[\beta^{\raisebox{0.5pt}{\tiny{$\,(1)$}}}_{kk}]\sin(2\phi_{k})\nonumber\\
&-&\Re[(\alpha^{\raisebox{0.5pt}{\tiny{$\,(1)$}}}_{kk})^2]\cos(2\phi_{k}))]\sinh^2r\big).
\end{eqnarray}

\subsubsection{Two-mode detection}
In this section we analyze the precision in the measurement of $\theta$ when two modes of the quantum field are detected. 
Our aim is again to find the second order contribution to the fidelity which will allow us to find the leading order term in the QFI and hence the lower bound on the error in estimation of the parameter $\theta$. 

Let us start analyzing the expansion of $\Lambda$, $\Gamma$ and $\Delta$ in the fidelity formula \eqref{FidelityMarianTwomodes}.
We first notice that 
\begin{eqnarray}\label{Lambdaexpan}
\Lambda(\theta,\theta+d\theta)&=&\prod_{i}(1-\nu^{2}_{i}(\theta))(1-\nu^{2}_{i}(\theta+d\theta))\nonumber\\
&=&(\nu_{1}^{\raisebox{0.5pt}{\tiny{$\,(1)$}}}\nu_{2}^{\raisebox{0.5pt}{\tiny{$\,(1)$}}}\theta(\theta+d\theta))^2+O(\theta^3(\theta+d\theta)^3))\nonumber\\
\end{eqnarray}
which implies that $\Lambda$ does not contribute to the term $\mathcal{C}^{(2)}$ in the fidelity expansion. Using the multiplicapitivity property of the determinants, we can write 
\begin{eqnarray}\label{gammaanddelta}
\Gamma &=&\text{det}(\mathds{1}+\sigma_{\oplus}(\theta)\mathbf{P}(\theta,\theta+d\theta))\nonumber\\
\Delta &=& \text{det}(\sigma_{\oplus}(\theta)+\mathbf{P}(\theta,\theta+d\theta))
\end{eqnarray}
 where $\mathbf{P}(\theta,\theta+d\theta):=s^{-1}(\theta) \sigma(\theta+d\theta)(s^{-1}(\theta))^{\dag}$.\\ 
For a given a matrix $\mathbf{M}$ with perturbative expansion 
\begin{eqnarray}
\mathbf{M}&=&\textbf{1}+ \mathbf{M}^{\raisebox{0.5pt}{\tiny{$\,(1,0)$}}}\theta+ \mathbf{M}^{\raisebox{0.5pt}{\tiny{$\,(0,1)$}}}d\theta+\mathbf{M}^{\raisebox{0.5pt}{\tiny{$\,(1,1)$}}}\theta d\theta
+\mathbf{M}^{\raisebox{0.5pt}{\tiny{$\,(2,0)$}}}\theta^2\nonumber\\
&+&\mathbf{M}^{\raisebox{0.5pt}{\tiny{$\,(0,2)$}}}d\theta^2,\nonumber
\end{eqnarray}
and eigenvlues 
\begin{eqnarray}
\lambda_{i}&=&1+\lambda_{i}^{{\raisebox{0.5pt}{\tiny{$\,(1,0)$}}}}\theta+\lambda_{i}^{{\raisebox{0.5pt}{\tiny{$\,(0,1)$}}}}d\theta+
\lambda_{i}^{{\raisebox{0.5pt}{\tiny{$\,(1,1)$}}}}\theta d\theta+
\lambda_{i}^{{\raisebox{0.5pt}{\tiny{$\,(2,0)$}}}}\theta^2\nonumber\\
&+&\lambda_{i}^{{\raisebox{0.5pt}{\tiny{$\,(0,2)$}}}}d\theta^2\nonumber\end{eqnarray}
we prove that
\begin{eqnarray}\label{determinantmexpansion}
\text{det}(\mathds{1}+\textbf{M})&=&16+8\text{Tr}[\mathbf{M}^{\raisebox{0.5pt}{\tiny{$\,(1,0)$}}}]\theta+8\text{Tr}[\mathbf{M}^{\raisebox{0.5pt}{\tiny{$\,(0,1)$}}}]d\theta\nonumber\\
&+&(4\sum_{i<j}\lambda_{i}^{\raisebox{0.5pt}{\tiny{$\,(1,0)$}}}\lambda_{i}^{\raisebox{0.5pt}{\tiny{$\,(0,1)$}}}+8\text{Tr}[\mathbf{M}^{\raisebox{0.5pt}{\tiny{$\,(1,1)$}}}])\theta d\theta\nonumber\\
&+&(4\sum_{i<j}\lambda_{i}^{\raisebox{0.5pt}{\tiny{$\,(1,0)$}}}\lambda_{i}^{\raisebox{0.5pt}{\tiny{$\,(1,0)$}}}+8\text{Tr}[\mathbf{M}^{\raisebox{0.5pt}{\tiny{$\,(2,0)$}}}])\theta^2\nonumber\\
&+&(4\sum_{i<j}\lambda_{i}^{\raisebox{0.5pt}{\tiny{$\,(0,1)$}}}\lambda_{i}^{\raisebox{0.5pt}{\tiny{$\,(0,1)$}}}+8\text{Tr}[\mathbf{M}^{\raisebox{0.5pt}{\tiny{$\,(0,2)$}}}])d\theta^2.\nonumber\\
\end{eqnarray}
Using \eqref{gammaanddelta} and \eqref{determinantmexpansion}  we find that,  up to third order corrections, $\Gamma= \Delta + \Delta_c$, where $\Delta_c$ is of the order $\theta(\theta + d\theta)$. Considering the perturbative expansions of $\Lambda$, $\Gamma$, $\Delta$ and the relation between $\Gamma$ and $\Delta$, we find that the term $\sqrt{(\sqrt{\Lambda}+\sqrt{\Gamma})^2-\Delta}$ in \eqref{two:mode:fidelity} does not have any term proportional to $d\theta^2$. Therefore, we conclude that the only term which contributes to the order $d\theta^2$ in fidelity is $\Gamma$. We can then obtain the second contribution to the Fidelity for a general two-mode Gaussian state as
\begin{eqnarray}\label{C22gen}
\mathcal{C}_{2}^{(2)}&=&
\frac{1}{16}\bigg[\big(\text{Tr}[(\sigma^{(0)})^{-1}\sigma^{(1)}]\big)^2
-\text{Tr}[((\sigma^{(0)})^{-1}\sigma^{(1)})^2]\nonumber\\
&+&4\text{Tr}[(\sigma^{(0)})^{-1}\sigma^{(2)}]\bigg].
\end{eqnarray}
Note that in order to find the second contribution $\mathcal{C}^{(2)}$ for any initial state of the quatum field we need to compute the covariance matrix elements up to second order.

Consider that the initial state of the quantum field is a generic Gaussian state for modes~$k$ and~$k^{\prime}$, and that all other modes are in their vacuum state.  The reduced covariance matrix for modes $k$ and~$k^{\prime}$ is given by
\be\label{initialcm}
\sigma_{kk^{\prime}}=\left(
               \begin{array}{cc}
                 \psi_k & \phi_{kk^{\prime}} \\
                 \phi^{T}_{kk^{\prime}} & \psi_k^{\prime}
               \end{array}
             \right)\,,
\ee
where $\psi_{k}$ and $\psi_{k^{\prime}}$ are the reduced covariance matrices of the modes $k$  and $k'$ respectively. The $2\times2$ matrix $\phi_{kk^{\prime}}$ contains the correlations between the two modes and it vanishes for product states. The transformed covariance matrix is then given by,
\be\label{transformedCM}
 \tilde{\sigma}_{kk^{\prime}}=\left(
                        \begin{array}{cc}
                          C_{kk} & C_{kk^{\prime}} \\
                          C_{k^{\prime}k} & C_{k^{\prime}k^{\prime}} \\
                        \end{array}
                      \right)\,,
\ee
where
\begin{align}
 C_{ij} &=\,\mathcal{M}_{ki}^T \psi _k \mathcal{M}_{kj}\,
    +\,\mathcal{M}_{k^{\prime}i}^{T}\phi^{T}_{kk^{\prime}}\mathcal{M}_{kj}\,
    +\,\mathcal{M}_{ki}^{T}\phi _{kk^{\prime}}\mathcal{M}_{k^{\prime}j}\nonumber\\
 &\ +\,\mathcal{M}_{k^{\prime}i}^{T}\psi_{k^{\prime}}\mathcal{M}_{k^{\prime}j}\,
    +\,\sum _{ n\neq k,k'}\mathcal{M}_{ni}^{T} \mathcal{M}_{nj}\,.
 \label{Cij}
\end{align}
We now assume that both modes~$k$ and~$k^{\prime}$ are in a displaced squeezed state, which is a product state. The $2\times2$ blocks in~(\ref{initialcm}) are $\psi_{k}=\psi_{k'}=\text{diag}\left(e^{r}\,,\,e^{-r}\right)$ while $\phi_{kk^{\prime}}$ vanishes. Using the equations (\ref{C22gen}) and \eqref{Cij} we find the contribution $\mathcal{C}^{(2)}$ to the fidelity as   
\begin{eqnarray}\label{Frankgenral}
\mathcal{C}_{2}^{(2)}&=&\frac{1}{4}\sum_{i,j}\Re\bigg[4 \cosh(r)(f^{i}_{\alpha}+f^{i}_{\beta})+ 4\sinh r G_{j}^{*2}\mathcal{G}^{\alpha \beta}_{jj}\nonumber\\
&+&2\cosh^2r(|\beta_{ij}^{\raisebox{0.5pt}{\tiny{$\,(1)$}}}|^2-f^{i}_{\alpha}+f^{i}_{\beta})-2\cosh^4r|\beta_{ij}^{\raisebox{0.5pt}{\tiny{$\,(1)$}}}|^2\nonumber\\
&-&2\sinh^2r(G_{j}^{*2}{\alpha_{ij}^{\raisebox{0.5pt}{\tiny{$\,(1)$}}}}^{2}+G_{j}^{2}{\beta_{ij}^{\raisebox{0.5pt}{\tiny{$\,(1)$}}}}^{2}+2G_{i}{\alpha_{ii}^{\raisebox{0.5pt}{\tiny{$\,(2)$}}}}^{*})\nonumber\\
&-&4\sinh2r(\alpha_{ij}^{\raisebox{0.5pt}{\tiny{$\,(1)$}}}\beta_{ij}^{\raisebox{0.5pt}{\tiny{$\,(1)$}}})
+2\sinh(2r)\cosh^2(r)\alpha_{ij}^{\raisebox{0.5pt}{\tiny{$\,(1)$}}}\beta_{ij}^{\raisebox{0.5pt}{\tiny{$\,(1)$}}}\nonumber\\
&+&\sinh^4r(|\alpha_{ij}^{\raisebox{0.5pt}{\tiny{$\,(1)$}}}|^2-|\beta_{ij}^{\raisebox{0.5pt}{\tiny{$\,(1)$}}}|^2-G_{j}^{*2}{\alpha_{ij}^{\raisebox{0.5pt}{\tiny{$\,(1)$}}}}^{2}
-G_{j}^{2}{\beta_{ij}^{\raisebox{0.5pt}{\tiny{$\,(1)$}}}}^{2})\nonumber\\
&-&\frac{1}{4}\sinh^22r(|\alpha_{ij}^{\raisebox{0.5pt}{\tiny{$\,(1)$}}}|^2-3|\beta_{ij}^{\raisebox{0.5pt}{\tiny{$\,(1)$}}}|^2-G_{j}^{*2}{\alpha_{ij}^{\raisebox{0.5pt}{\tiny{$\,(1)$}}}}^{2}\nonumber\\
&-&G_{j}^{2}{\beta_{ij}^{\raisebox{0.5pt}{\tiny{$\,(1)$}}}}^{2})\bigg]
\end{eqnarray}
where we have defined
\begin{align*}
f_{\alpha}^{i}:=&\,\frac{1}{2}\sum_{n\neq k,k'}|\alpha_{ni}^{\raisebox{0.5pt}{\tiny{$\,(1)$}}}|^2\\
f_{\beta}^{i}:=&\, \frac{1}{2}\sum_{n\neq k,k'}|\beta_{ni}^{\raisebox{0.5pt}{\tiny{$\,(1)$}}}|^2\\
\mathcal{G}^{\alpha\beta}_{ij}:=&\,\sum_{n\neq k,k'} \alpha_{ni}^{\raisebox{0.5pt}{\tiny{$\,(1)$}}}{\beta_{nj}^{\raisebox{0.5pt}{\tiny{$\,(1)$}}}}^{*}.
\end{align*}

The first moments of the initial state in this case can be written as $\langle X_0\rangle=(\sqrt{2}\delta,0,\sqrt{2}\delta,0)^{T}$. Again after computing the first order contribution to the first moments of the quantum field after the Bogoliubov transformation we find the second order contribution $\mathcal{E}^{\raisebox{0.5pt}{\tiny{$\,(2)$}}}$ as 
\begin{eqnarray}\label{FrankDisplaced}
 \mathcal{E}_{2}^{\raisebox{0.5pt}{\tiny{$\,(2)$}}}&=&2\delta^2\big[\cosh r(|A_{kk'}|^{2}+|A_{k'k}|^{2})\nonumber\\
 &+&\sinh r (\cos(2\phi_{k})|A_{kk'}|^{2}+\sin(2\phi_{k})\Im[A_{kk'}^{2}]\nonumber\\
 &+&\cos(2\phi_{k'})|A_{k'k}|^{2}+\sin(2\phi_{k'})\Im[A_{k'k}^{2}]\big], 
 \end{eqnarray}
 where $A_{ij}=\alpha_{ii}^{\raisebox{0.5pt}{\tiny{$\,(1)$}}}+\alpha_{ij}^{\raisebox{0.5pt}{\tiny{$\,(1)$}}}-\beta_{ii}^{\raisebox{0.5pt}{\tiny{$\,(1)$}}}-\beta_{ij}^{\raisebox{0.5pt}{\tiny{$\,(1)$}}}$. Note that this term depends on both the amount of initial dispalcement $\delta$ and the amount of initial squeezing $r$ in each mode. Also it is worth pointing out that having non-zero initial displacement is necessary for the term $\mathcal{E}^{\raisebox{0.5pt}{\tiny{$\,(2)$}}}$ to contribute to the QFI.

\section{Example: Cavity in non-uniform motion}
In this section we show how to apply our techniques to estimate acceleration using a massless scalar field $\Phi$ confined within rigid moving boundary conditions. In this case, a massless scalar field is a good approximation to one of the polarisation modes of the electromagnetic field \cite{Friis:Lee:Louko:13}. We consider the scenario where a cavity of length $L$ is initially inertial in a $(1+1)$-dimensional flat spacetime with co-ordinates $(t,x)$ and then moves non-inertially for a finite period of time. Initially, we will consider general trajectories and towards the end of the example we will present results for a specific trajectory. Independently of the trajectory details, we consider that at all times the field vanishes at the cavity walls, i.e. $\Phi_{n}(t,x_L)=\Phi_{n}(t,x_R)=0$. The field $\Phi$ satisfies the Klein-Gordon equation $\square\Phi=0$, where the d'Alambertian takes the simple form $\square=\partial_t^2-c^2\partial_x^2$ in Minkowski coordinates. The solutions to the Klein-Gordon equation are plane waves. After imposing the corresponding boundary conditions we find the mode solutions $\phi_n$. We assume that, after the cavity has undergone non-inertial motion, the cavity will once more move inertially and it is then possible to find a new set of field modes $\tilde{\phi}_n$.  The two sets of modes $\phi_n$  and $\tilde{\phi}_n$ are related by a Bogoliubov transformation, details of this transformation can be found in \cite{alphacentauri}. The transformation can be expanded in the form (\ref{Maclaurin expansion}) where the small parameter in this case is $h=\frac{aL}{c^2}$. Here $a$ is the proper acceleration at the center of the cavity and $c$ is the speed of light. In this particular setup, the diagonal first order Bogoliubov coefficients are zero, i.e. $\alpha^{\raisebox{0.5pt}{\tiny{$\,(1)$}}}_{ii}=\beta^{\raisebox{0.5pt}{\tiny{$\,(1)$}}}_{ii}=0$. Our aim is to employ the formulas presented in the sections above to estimate the proper acceleration $a$. To do so we consider different initial states of the cavity and we compare the quantum Fisher information between these different cases. In \cite{RQM} this example was studied only in the case that the initial state of the field was a state with two modes, each one of them squeezed. Now we will be able to analyse what initial states provide a better estimation of the acceleration.

\subsection{Single-mode detection}
Let us start with the case where all the modes of the quantum field inside the cavity are in their vacuum state except mode $k$ which is in a squeezed displaced vacuum state.  Since in the case of a moving cavity the diagonal first order contribution to the Bogoliubov coefficients vanish, the term $\mathcal{E}_{1}^{\raisebox{0.5pt}{\tiny{$\,(2)$}}}$ is zero, which implies that there is no advantage in measuring $h$ by initially displacing the state.
 Using equations \eqref{QFI} and \eqref{QFIsinglemodegenral} we find that the leading order term of the QFI is, \begin{equation}
H_{1}=4\left[(f_{\alpha}^{k}+f_{\beta}^{k})\cosh(r)-\Re[\alpha^{\raisebox{0.5pt}{\tiny{$\,(1)$}}}_{nk}{\beta^{\raisebox{0.5pt}{\tiny{$\,(1)$}}}_{nk}}^{*}] \sinh(r)\right],
\end{equation}  
which only depends on the first order Bogoliubov coefficients and the squeezing parameter.  The Bogoliubov coefficients here correspond to any trajectory followed by the cavity. We note that when $r=0$ then $H_{1}$ is very small. This will give rise to a very large error which can be decreased by increasing the squeezing in the state. As already noted, initial displacement will not affect the precision.

\subsection{Two modes in a product form} 
Let us consider a different initial state for the field inside the cavity. We assume that two modes $k$ and $k'$ are initially prepared in a separable state where each mode is in a displaced squeezed vacuum. To simplify calculations we  assume the same amount of squeezing $r$ and displacement $\delta$ in each mode. Using equations \eqref{Frankgenral} and \eqref{FrankDisplaced} and considering that  $\alpha^{\raisebox{0.5pt}{\tiny{$\,(1)$}}}_{nn}=\beta^{\raisebox{0.5pt}{\tiny{$\,(1)$}}}_{nn}=0$ we find, 
\begin{eqnarray}\label{Frankcavity}
\mathcal{C}_{2}^{(2)}&=&\frac{1}{4}\Re\bigg[4 \cosh r(f^{k}_{\alpha}+f^{k}_{\beta}+f^{k'}_{\alpha}+f^{k'}_{\beta})-4\cosh^4r|\beta_{kk'}^{\raisebox{0.5pt}{\tiny{$\,(1)$}}}|^2\nonumber\\
&+& 4\cosh^2r(2|\beta_{kk'}^{\raisebox{0.5pt}{\tiny{$\,(1)$}}}|^2+f^{k}_{\beta}-f^{k}_{\alpha}+f^{k'}_{\beta}-f^{k'}_{\alpha})
\nonumber\\
&-&4\sinh^2r(G_{k'}^{*2}{\alpha_{kk'}^{\raisebox{0.5pt}{\tiny{$\,(1)$}}}}^{2}
+G_{k'}^{2}{\beta_{kk'}^{\raisebox{0.5pt}{\tiny{$\,(1)$}}}}^{2}+G_{k}^{*}{\alpha_{kk}^{\raisebox{0.5pt}{\tiny{$\,(2)$}}}}+G_{k'}^{*}{\alpha_{k'k'}^{\raisebox{0.5pt}{\tiny{$\,(2)$}}}})\nonumber\\
&-&4\sinh2r(\alpha_{kk'}^{\raisebox{0.5pt}{\tiny{$\,(1)$}}}\beta_{kk'}^{\raisebox{0.5pt}{\tiny{$\,(1)$}}}+\alpha_{k'k}^{\raisebox{0.5pt}{\tiny{$\,(1)$}}}\beta_{k'k}^{\raisebox{0.5pt}{\tiny{$\,(1)$}}})\nonumber\\
&+& 4\sinh r(G_{k}^{*2}\mathcal{G}^{\alpha \beta}_{kk}+G_{k'}^{*2}\mathcal{G}^{\alpha \beta}_{k'k'})\nonumber\\
&+&4\sinh2r\cosh^2r(\alpha_{kk'}^{\raisebox{0.5pt}{\tiny{$\,(1)$}}}\beta_{kk'}^{\raisebox{0.5pt}{\tiny{$\,(1)$}}}+\alpha_{k'k}^{\raisebox{0.5pt}{\tiny{$\,(1)$}}}\beta_{k'k}^{\raisebox{0.5pt}{\tiny{$\,(1)$}}})\nonumber\\
&+&2\sinh^4r(|\alpha_{kk'}^{\raisebox{0.5pt}{\tiny{$\,(1)$}}}|^2-|\beta_{kk'}^{\raisebox{0.5pt}{\tiny{$\,(1)$}}}|^2-{G_{k'}^{*}}^{2}{\alpha_{kk'}^{\raisebox{0.5pt}{\tiny{$\,(1)$}}}}^{2}-G_{k'}^{2}{\beta_{kk'}^{\raisebox{0.5pt}{\tiny{$\,(1)$}}}}^{2})\nonumber\\
&-&\frac{1}{2}\sinh^22r(|\alpha_{kk'}^{\raisebox{0.5pt}{\tiny{$\,(1)$}}}|^2-3|\beta_{kk'}^{\raisebox{0.5pt}{\tiny{$\,(1)$}}}|^2
-{G_{k'}^{*}}^{2}{\alpha_{kk'}^{\raisebox{0.5pt}{\tiny{$\,(1)$}}}}^{2}\nonumber\\
&-&G_{k'}^{2}{\beta_{kk'}^{\raisebox{0.5pt}{\tiny{$\,(1)$}}}}^{2})\bigg].
\end{eqnarray}
and
\begin{eqnarray}\label{FrankDisplaced}
 \mathcal{E}_{2}^{\raisebox{0.5pt}{\tiny{$\,(2)$}}}&=&2\delta^2\big[\cosh r(|\alpha_{kk'}^{\raisebox{0.5pt}{\tiny{$\,(1)$}}}-\beta_{kk'}^{\raisebox{0.5pt}{\tiny{$\,(1)$}}}|^{2}+|\alpha_{k'k}^{\raisebox{0.5pt}{\tiny{$\,(1)$}}}-\beta_{k'k}^{\raisebox{0.5pt}{\tiny{$\,(1)$}}}|^{2})\nonumber\\
 &+&\sinh r (\cos(2\phi_{k})|\alpha_{kk'}^{\raisebox{0.5pt}{\tiny{$\,(1)$}}}-\beta_{kk'}^{\raisebox{0.5pt}{\tiny{$\,(1)$}}}|^{2}\nonumber\\
 &+&\sin(2\phi_{k})\Im[(\alpha_{kk'}^{\raisebox{0.5pt}{\tiny{$\,(1)$}}}-\beta_{kk'}^{\raisebox{0.5pt}{\tiny{$\,(1)$}}})^{2}]\nonumber\\
 &+&\cos(2\phi_{k'})|\alpha_{k'k}^{\raisebox{0.5pt}{\tiny{$\,(1)$}}}-\beta_{k'k}^{\raisebox{0.5pt}{\tiny{$\,(1)$}}}|^{2}\nonumber\\
 &+&\sin(2\phi_{k'})\Im[(\alpha_{k'k}^{\raisebox{0.5pt}{\tiny{$\,(1)$}}}-\beta_{k'k}^{\raisebox{0.5pt}{\tiny{$\,(1)$}}})^{2}]\big], 
 \end{eqnarray}
 The QFI is this case in given by $H_2=4(\mathcal{C}^{(2)}_2+\mathcal{E}^{(2)}_2)$.
\begin{figure}[t]
\includegraphics[width=\linewidth]{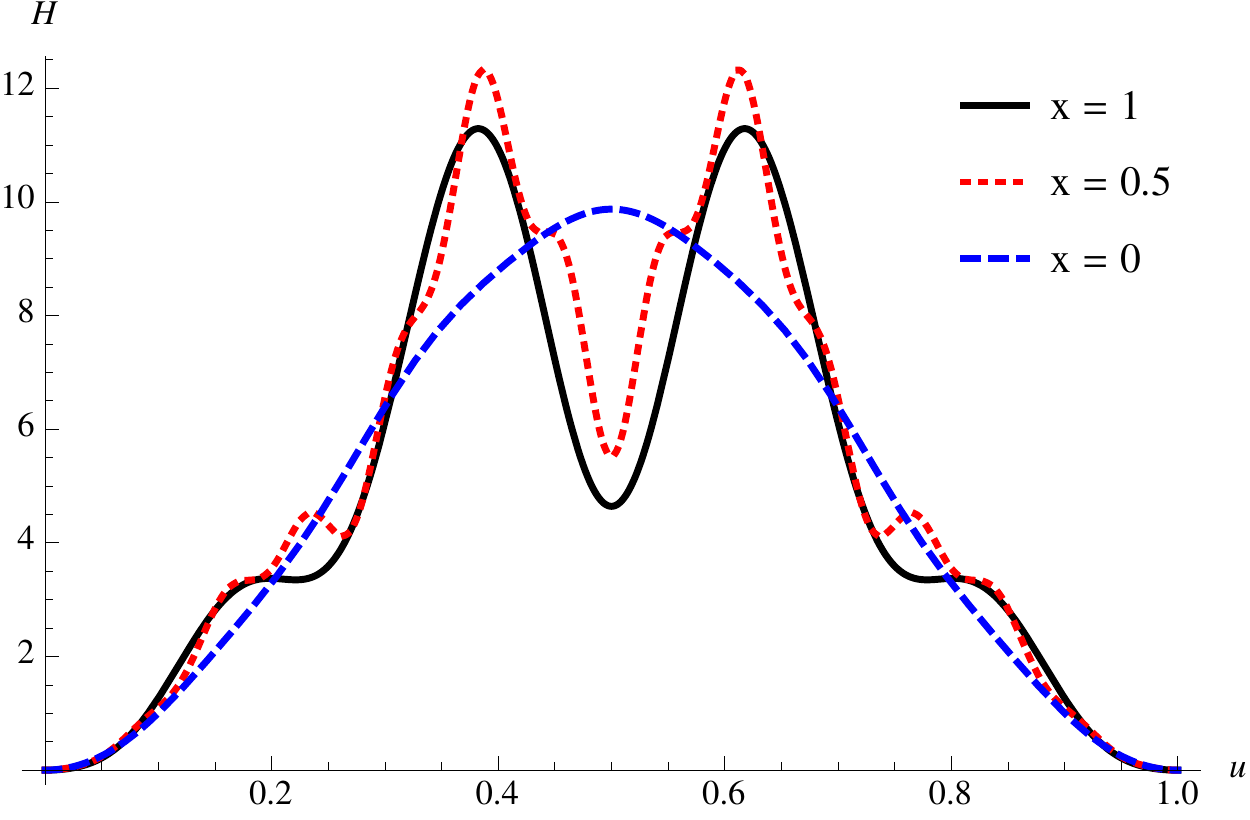}
\caption{Quantum Fisher information vs. $u=\frac{h\tau}{4L\arctanh(h/2)}$ for modes $k=1$ and $k'=2$ when the initial state is only squeezed $x=0$, only displaced $x=1$ and with an equal amount of energy associated to squeezing and displacement $x=0.5$.} \label{Frankxcomparison}
\end{figure}

\subsection{Two-mode squeezed state}
Finally we consider an initial state for two modes $k,k'$ containing entanglement. The state is known as the two-mode squeezed state and is given by $\phi_{kk}=\phi_{k'k'}=\cosh(r)\mathds{1}$ and $\phi_{kk'}=\sinh(r)\sigma_z$ where $\sigma_z$ is the Pauli z-matrix. A two mode squeezed state has zero first moments, therefore, the contribution to the fidelity coming form the exponential term $\mathcal{E}^{(2)}_3$ vanishes. Using once more equation \eqref{Frankgenral} we find that $H_3=4\mathcal{C}_{3}^{(2)}$ which in terms of Bogoliubov coefficients yields,

\begin{eqnarray}\label{H3}
H_3&=&\Re\bigg[4 \cosh r(f^{k}_{\alpha}+f^{k}_{\beta}+f^{k'}_{\alpha}+f^{k'}_{\beta})-4\cosh^4r|\beta_{kk'}^{\raisebox{0.5pt}{\tiny{$\,(1)$}}}|^2\nonumber\\
&+& 4\cosh^2r(2|\beta_{kk'}^{\raisebox{0.5pt}{\tiny{$\,(1)$}}}|^2+f^{k}_{\beta}-f^{k}_{\alpha}+f^{k'}_{\beta}-f^{k'}_{\alpha})
\nonumber\\
&-&4\sinh^2r(-\cos^2(\phi_k+\phi_k'){\alpha_{kk'}^{\raisebox{0.5pt}{\tiny{$\,(1)$}}}}^{2}
+G_{k'}^{2}{\beta_{kk'}^{\raisebox{0.5pt}{\tiny{$\,(1)$}}}}^{2}+G_{k}^{*}{\alpha_{kk}^{\raisebox{0.5pt}{\tiny{$\,(2)$}}}}\nonumber\\
&+&G_{k'}^{*}{\alpha_{k'k'}^{\raisebox{0.5pt}{\tiny{$\,(2)$}}}}
-\Im[G_{k}G_{k'}\alpha_{kk'}^{\raisebox{0.5pt}{\tiny{$\,(1)$}}}]\Re[\alpha_{kk'}^{\raisebox{0.5pt}{\tiny{$\,(1)$}}}]\sin(\phi_{k}+\phi_{k'}))\nonumber\\
&+&4\sinh r (\Re[\mathcal{G}^{\alpha \beta}_{kk'}+\mathcal{G}^{\alpha \beta}_{k'k}]\cos[\phi_k+\phi_{k'}]\nonumber\\
&-&\Im[\mathcal{G}^{\alpha \beta}_{k'k}+\mathcal{G}^{\alpha \alpha}_{kk'}]\sin(\phi_{k}+\phi_{k'}))\nonumber\\
&-&\frac{1}{2}\sinh^22r(2|\alpha_{kk'}^{\raisebox{0.5pt}{\tiny{$\,(1)$}}}|^2-3|\beta_{kk'}^{\raisebox{0.5pt}{\tiny{$\,(1)$}}}|^2
-{G_{k'}}^{2}{\beta_{kk'}^{\raisebox{0.5pt}{\tiny{$\,(1)$}}}}^{2}\bigg].\nonumber
\end{eqnarray}
Note that in the limit of zero squeezing for one-mode detection scheme the QFI reduces to
\begin{equation}\label{vacuum1}
H_{1}\big|_{r=0}=8f_{\beta}^{k},
\end{equation}  
whereas for the two-mode detection scheme we find the QFI as
\begin{eqnarray}\label{vacuum2}
H_2\big|_{r=0}=H_3\big|_{r=0}=4\mathcal{C}_{vac}^{(2)}&=&8f^{k}_{\beta}+8f^{k'}_{\beta}+4|\beta_{kk'}^{\raisebox{0.5pt}{\tiny{$\,(1)$}}}|^2.\nonumber
\end{eqnarray}
We note that the term $|\beta_{kk'}^{\raisebox{0.5pt}{\tiny{$\,(1)$}}}|^2$ is directly related to the entangelement generated between the modes $k$ and $k'$ due to the non-uniform motion of the cavity. It was shown that the negativity $\mathcal{N}$ (which quantifies entanglement) generated between the modes after the Bogoliubov transformation has the simple expression  $\mathcal{N}=|\beta_{kk'}^{\raisebox{0.5pt}{\tiny{$\,(1)$}}}|$ \cite{vacuumentanglement}. From equations \eqref{vacuum1} and \eqref{vacuum2} one can see that $H_2\big|_{r=0}=H_3\big|_{r=0}\sim 2\,H_{1}\big|_{r=0}+4\mathcal{N}^2\geq H_{1}\big|_{r=0}$. Therefore, employing two modes provides a higher precision compreard to the single mode case due to the entanglement generated by the transformation.\\

\begin{figure}[t]
\includegraphics[width=\linewidth]{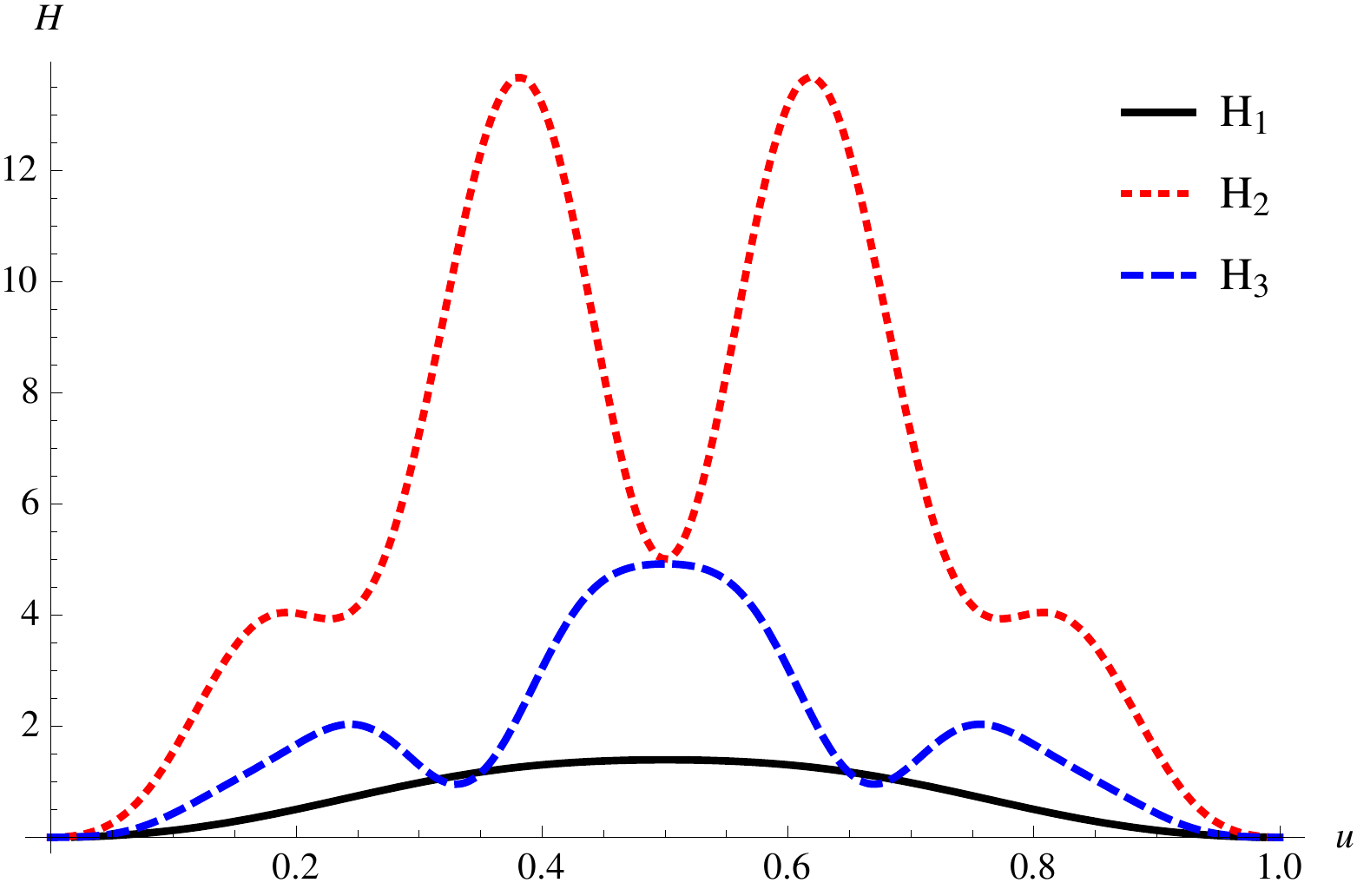}
\caption{Quantum Fisher information vs. $u=\frac{h\tau}{4L\arctanh(h/2)}$ for three different initial states of the cavity with the same average energy. A single-mode squeezed state in mode $k=1$ (solid black line),  two single-mode squeezed states in a product form in modes $k=1$ and $k'=2$ (dotted red line) and a two-mode squeezed state in modes $k=1$ and $k'=2$ (dashed blue line).} \label{fig:results}
\end{figure}

So far we have presented our results considering that the cavity follows an arbitrary trajectory. In order to make a comparison on the estimation of acceleration using the states analysed above, we specify our example further. We will consider that the cavity is initially inertial and then undergoes a single period of uniform acceleration $a$ and duration $\tau$  after which the cavity returns to inertial motion. In this case the Bogoliubov coefficients $\alpha,\beta$ are given in terms of simple expressions of the inertial-to-Rindler coefficients ${}_{\circ}\alpha,{}_{\circ}\beta$. These are the coefficients that relate the solutions of the Klein-Gordon equation in the inertial frame to the solutions in the uniformly accelerated frame. For the particular travel scenario that we consider here,  the final Bogoliubov coefficients are given by 
\begin{eqnarray}\label{BBBBogos}
\alpha^{\raisebox{0.5pt}{\tiny{$\,(1)$}}}_{ij}&=&{}_{\circ}\alpha^{\raisebox{0.5pt}{\tiny{$\,(1)$}}}_{ij}(G_i-G_{j}),\nonumber\\
\beta^{\raisebox{0.5pt}{\tiny{$\,(1)$}}}_{ij}&=&{}_{\circ}\beta^{\raisebox{0.5pt}{\tiny{$\,(1)$}}}_{ij}(G_i-G^{*}_{j}),\nonumber\\
\alpha^{\raisebox{0.5pt}{\tiny{$\,(2)$}}}_{ij}&=&G_i {}_{\circ}\alpha^{\raisebox{0.5pt}{\tiny{$\,(2)$}}}_{ij}-G_j {}_{\circ}\alpha^{\raisebox{0.5pt}{\tiny{$\,(2)$}}}_{ji}+\sum_{k}[G_{k}{}_{\circ}\alpha^{\raisebox{0.5pt}{\tiny{$\,(1)$}}}_{ki}{}_{\circ}\alpha^{\raisebox{0.5pt}{\tiny{$\,(1)$}}}_{kj}\nonumber\\
&-&G^{*}_{k}{}_{\circ}\beta^{\raisebox{0.5pt}{\tiny{$\,(1)$}}}_{ki}{}_{\circ}\beta^{\raisebox{0.5pt}{\tiny{$\,(1)$}}}_{kj}],\nonumber\\
\alpha^{\raisebox{0.5pt}{\tiny{$\,(2)$}}}_{ij}&=&G_i {}_{\circ}\beta^{\raisebox{0.5pt}{\tiny{$\,(2)$}}}_{ij}-G_j^{*} {}_{\circ}\beta^{\raisebox{0.5pt}{\tiny{$\,(2)$}}}_{ji}+\sum_{k}[G_{k}{}_{\circ}\alpha^{\raisebox{0.5pt}{\tiny{$\,(1)$}}}_{ki}{}_{\circ}\beta^{\raisebox{0.5pt}{\tiny{$\,(1)$}}}_{kj}\nonumber\\
&-&G^{*}_{k}{}_{\circ}\beta^{\raisebox{0.5pt}{\tiny{$\,(1)$}}}_{ki}{}_{\circ}\alpha^{\raisebox{0.5pt}{\tiny{$\,(1)$}}}_{kj}].\nonumber\\
\end{eqnarray}
where ${}_{\circ}\alpha^{\raisebox{0.5pt}{\tiny{$\,(n)$}}}_{ij}$ and ${}_{\circ}\beta^{\raisebox{0.5pt}{\tiny{$\,(n)$}}}_{ij}$ are the $n$-th order contributions of the Bogoliubov coefficients ${}_{\circ}\alpha,{}_{\circ}\beta$ which are given in full detail in \cite{alphacentauri}. The coefficients $G_{n}=\exp[i\tilde{\omega}_n \tau]$ are phase factors that each mode $n$ accumulates due to free evolution during the acceleration period. The angular frequencies with respect to the proper time at the center of the cavity are given by $\tilde{\omega}_n=\frac{n\pi h}{2L\arctanh(h/2)}$. We choose to plot our results in terms of the dimensionless parameter $u=\frac{\tilde{\omega}_n\tau}{2\pi n}=\frac{h\tau}{4L\arctanh(h/2)}$. 

In order to make a fair comparison between the schemes considering different initial states, we assume that the energy of the initial state is the same in each case, i.e. 
$E_0=\hbar N_k(\omega_k + \omega_{k'})$, where $\omega_n=\frac{n\,\pi\,c}{L}$ is the frequency of mode $n$ and $N_{n}=\sinh^2r+\delta^2$ is the average photon number of mode $n$ before the Bogoliubov transformation. We define the parameter $x$  to denote the fraction of initial energy associated to squeezing i.e., $\sinh^2 r= x N_{n}$  and therefore the fraction of energy associated to displacement is given by $\delta^2=(1-x)N_n$. In figure (\ref{Frankxcomparison}) we plot the QFI $H_2$ as a function of $u$ for $x=0,0.5$ and $1$.

In order to understand in more detail the role of squeezing in the analysis, we plot our results in the case that the initial displacements are zero.
In figure (\ref{fig:results}) we compare the quantum Fisher information as a function of $u$ for the three different initial states: Single-mode squeezed state (solid black), a product state of two single-mode squeezed states (dotted red) and an entangled two-mode squeezed state (dashed blue). We have used different squeezing parameters for different initial states in such a way that the total energy of the states is the same.  The two single-mode squeezed state provides us with a better precision in the estimation of proper acceleration for every value of $u$. This improvement is due to the entanglement generated between the modes $k$ and $k'$. Furthermore, the entanglement generated due to the motion of the cavity reaches a local minimum as shown in \cite{nicoivy}. This has a direct consequence on the QFI, which also has a minumum for $u=0.5$, as can be seen in figure (\ref{fig:results}).
 It is worth pointing out that the comparison between the performance of a single-mode squeezed state and the performance of a two-mode squeezed state, i.e. $H_{1}$ and $H_{3}$, needs to be done at a particular proper time $\tau$. In other words, for some observers the single-mode squeezed state is a better choice of initial state while for the others it is more convinient to prepare the two modes $k$ and $k'$ in a two-mode squeezed state. 
\section{Conclusions}

In this paper we have provided techniques for the optimal estimation of parameters which appear in quantum field theory in curved spacetime. This enables the estimation of parameters such as proper accelerations, proper times, relative distances, gravitational field strenghts, as well as spacetime parameters of interest such as the expansion rate of the Universe or the mass of a black hole. We have combined techniques from quantum metrology, continuous variable quantum information, symplectic geometry and quantum field theory in order to develop a framework which is applicable in a generic scenario in which the parameter to be estimated is encoded in a Bogoliubov transformation. By resticting the analysis to small parameters and Gaussian states, we are able to provide analytical formulas for the QFI in terms of Bogoliubov coefficients. 

We apply our results to a scenario of current physical interest, namely the estimation of the acceleration of a cavity which undergoes non-uniform motion. This example was studied before in \cite{RQM} considering an initial state consisting of two single mode squeezed states in a product form. Here we extend the analysis to one-mode and two-mode Gaussian states. We show that the generation of entanglement in two-mode detection schemes increases the precision, therefore, improving single mode schemes. 

Our techniques are applicable to analogue gravity systems where the effects of space-time on quantum fields can be investigated in realisable experimental setups \cite{Visser:05}.

\section{acknowledgements}
We thank Carlos Sab\'in, Antony Lee, Jandu Dradouma and Nicolai Friis for useful discussions and comments. M.~A. and I.~F. acknowledge support from EPSRC (CAF Grant No.~EP/G00496X/2 to I.~F.). D.~E.~B.  was supported by the UK Engineering and Physical Science Research Council grant number EP/J005762/1. D.~E.~B. also acknowledges hospitality from the University of Nottingham. 


\begin{thebibliography}{100}
\addcontentsline{toc}{section}{References}


\bibitem{casimirwilson}
C.~M.~Wilson, G.~Johansson, A.~Pourkabirian, M.~Simoen, J.~R.~Johansson, T.~Duty, F.~Nori and P.~Delsing,\
\href{http://dx.doi.org/10.1038/nature10561}{Nature\ (London)\ \textbf{479}, 376 
(2011)}.

\bibitem{zeilingerteleport}
X.-S.~Ma, T.~Herbst, T.~Scheidl, D.~Wang, S.~Kropatschek, W.~Naylor, A.~Mech, B.~Wittmann, J.~Kofler, E.~Anisimova, V.~Makarov, T.~Jennewein, R.~Ursin, and A.~Zeilinger,\
\href{http://dx.doi.org/10.1038/nature11472}{Nature (London)\ \textbf{489}, 269}\\ 
\href{http://dx.doi.org/10.1038/nature11472}{(2012)}.

\bibitem{SchillerEtalSpaceOpticalClock2012}
S.~Schiller, A.~G{\"o}rlitz, A.~Nevsky, S.~Alighanbari, S.~Vasilyev, C.~Abou-Jaoudeh, G.~Mura, T.~Franzen, U.~Sterr, S.~Falke, et al.,\
e-print \href{http://arxiv.org/abs/1206.3765}{arXiv:1206.3765} [quant-ph] (2012).

\bibitem{rideout}
D.~Rideout, T.~Jennewein, G.~Amelino-Camelia, T.~F.~Demarie, B.~L.~Higgins, A.~Kempf, A.~Kent, R.~Laflamme, X.~Ma, R.~B.~Mann, E.~Mart\'{i}n-Mart\'{i}nez, N.~C.~Menicucci, J.~Moffat, C.~Simon, R.~Sorkin, L.~Smolin, and D.~R.~Terno,\
\href{http://dx.doi.org/10.1088/0264-9381/29/22/224011}{Class.\ Quantum\ Grav.}\
\href{http://dx.doi.org/10.1088/0264-9381/29/22/224011}{\textbf{29}, 
224011 (2012)}
.

\bibitem{aspachs}
M.~Aspachs, G.~Adesso, and I.~Fuentes,\
\href{http://dx.doi.org/10.1103/PhysRevLett.105.151301}{Phys.\ Rev.}\
\href{http://dx.doi.org/10.1103/PhysRevLett.105.151301}{Lett.\ \textbf{105}, 
151301 (2010)}.

\bibitem{downes}
T.~G.~Downes, G.~J.~Milburn, and C.~M.~Caves,\
e-print \href{http://arxiv.org/abs/1108.5220}{arXiv:1108.5220} [gr-qc] (2012).

\bibitem{RQM}
M.~Ahmadi, D.~E.~Bruschi, N.~Friis, C.~Sabn, G.~Adesso, and I.~Fuentes,\
e-print \href{http://arxiv.org/abs/1307.7082}{arXiv:1307.7082} [quant-ph] (2013).

\bibitem{fredericivy}
J.~L.~Ball, I.~Fuentes-Schuller, and F.~P.~Schuller,\
\href{http://dx.doi.org/10.1016/j.physleta.2006.07.028}{Phys.}\
\href{http://dx.doi.org/10.1016/j.physleta.2006.07.028}{Lett.\ A \textbf{359}, 550 
(2006)}.

\bibitem{HoslerKok2013}
D.~J.~Hosler and P.~Kok,\
e-print \href{http://arxiv.org/abs/1306.3144}{arXiv:1306.3144} [quant-ph] (2013).

\bibitem{SaleckerWigner}
H.~Salecker, and E.~P.~Wigner,\
\href{http://link.aps.org/doi/10.1103/PhysRev.109.571}{Phys.\ Rev.\ \textbf{109}, 571--577,2}\\ 
\href{http://link.aps.org/doi/10.1103/PhysRev.109.571}{(1958)}.

\bibitem{footnote}
The expansion is only possible when positive and negative modes solutions can be distinguished, i.e. when the space-time admits a time-like Killing vector field. In the case of a continuous mode decomposition, it is possible to construct a discrete basis formed of wave packets of continuous modes. 

\bibitem{alphacentauri}
D.~E.~Bruschi, I.~Fuentes, and J.~Louko,\
\href{http://dx.doi.org/10.1103/PhysRevD.85.061701}{Phys.\ Rev.\ D}\ \href{http://dx.doi.org/10.1103/PhysRevD.85.061701}{\textbf{85}, 
 061701(R) (2012)}.

\bibitem{ourreview}
 G.~Adesso and F.~Illuminati, \href{http://dx.doi.org/10.1088/1751-8113/40/28/S01}{J. Phys. A: Math. Theor.}\\ \href{http://dx.doi.org/10.1088/1751-8113/40/28/S01}{ {\bf 40} 7821, (2007)}.


\bibitem{nicoivy}
N.~Friis and I.~Fuentes,\
\href{http://dx.doi.org/10.1080/09500340.2012.712725}{J.\ Mod.\ Opt.\ \textbf{60}, 
22 
(2013)}.

\bibitem{BirrellDavies:QFbook}
N.~D.~Birrell and P.~C.~W.~Davies, \textit{Quantum Fields in Curved Space\/} (Cambridge University Press, Cambridge, England, 1982).

\bibitem{Friis:Lee:Louko:13}
N.~Friis, A. R. Lee and J.~Louko, \href{http://dx.doi.org/10.1103/PhysRevD.88.064028}{Phys.}\
\href{http://dx.doi.org/10.1103/PhysRevD.88.064028}{Rev.\ D\ \textbf{88}, 
 064028 (2013)}.

\bibitem{givyericsson}
G.~Adesso, I.~Fuentes-Schuller, and M.~Ericsson,\
\href{http://dx.doi.org/10.1103/PhysRevA.76.062112}{Phys.}\
\href{http://dx.doi.org/10.1103/PhysRevA.76.062112}{Rev.\ A\ \textbf{76}, 
 062112 (2007)}.

\bibitem{gcqg}
G.~Adesso, S.~Ragy, and D.~Girolami, 
\href{http://dx.doi.org/10.1088/0264-9381/29/22/224002}{Class. Quantum}\\ \href{http://dx.doi.org/10.1088/0264-9381/29/22/224002} { Grav. {\bf 29},  224002  (2012)}.

\bibitem{advances}
V.~Giovanetti, S.~Lloyd, and L.~Maccone,\
\href{http://dx.doi.org/10.1038/nphoton.2011.35}{Nature\ Photon.\ \textbf{5},} \href{http://dx.doi.org/10.1038/nphoton.2011.35}{222 
(2011)}.

\bibitem{paris}
M.~ G.~A.~Paris,\
\href{http://www.worldscientific.com/doi/abs/10.1142/S0219749909004839}{Int. J. Quant. Inf.\ \textbf{7}, 125 (2009)}.

\bibitem{Cramer:Methods1946}
H.~Cram$\mathrm{\acute{e}}$r,\
\textit{Mathematical Methods of Statistics}\
(Princeton University, Princeton, NJ, 1946).

\bibitem{BraunsteinCaves1994}
S.~L.~Braunstein and C.~M.~Caves,\
\href{http://dx.doi.org/10.1103/PhysRevLett.72.3439}{Phys.\ Rev.\ Lett.}\\ \href{http://dx.doi.org/10.1103/PhysRevLett.72.3439} {\textbf{72}, 
 3439 
 (1994)}.
 
 
\bibitem{monras}
A. Monras, e-print \href{http://arxiv.org/abs/1303.3682}{arXiv:1303.3682} [quant-ph] (2013). 

\bibitem{MarianMarian}
P.~Marian and T.~A.~Marian,\
\href{http://dx.doi.org/10.1103/PhysRevA.86.022340}{Phys.\ Rev.\ A \textbf{86}, 
022340} \href{http://dx.doi.org/10.1103/PhysRevA.86.022340}{(2012)}.
  
  
\bibitem{Weedbrook} C.~Weedbrook, S.~Pirandola, N.~J.~Cerf, T.~C.~Ralph, J.~H.~ Shapiro and S.~Lloyd,\ 
\href{http://rmp.aps.org/abstract/RMP/v84/i2/p621_1}{Rev. Mod. Phys. \textbf{49}}
\href{http://rmp.aps.org/abstract/RMP/v84/i2/p621_1}{(2012)}. 
 
\bibitem{BraunsteinCaves}
S.~L.~Braunstein and C.~M.~Caves,\
\href{http://prl.aps.org/abstract/PRL/v72/i22/p3439_1}{Phys.\ Rev.\ Lett. \textbf{72},
3439–3443} \href{http://prl.aps.org/abstract/PRL/v72/i22/p3439_1}{(1994)}. 
  
\bibitem{vacuumentanglement}
N.~Friis, D.~E.~Bruschi, J.~Louko and I.~Fuentes,\
\href{http://prd.aps.org/abstract/PRD/v85/i8/e081701}{Phys.\ Rev.\ D\ \textbf{85}, 
 081701(R) 
(2012)}. 
 
\bibitem{Visser:05}
C.~Barcel\'o, S.~Liberati and M.~Visser,\
\href{http://www.livingreviews.org/lrr-2005-12}{Living Reviews in Relativity\ \textbf{12} (8), 
(2005)}. 

%
%
  









































.
































%















\end{thebibliography}

\end{document}